\documentclass   {article}
\def\bbbone{{\mathchoice {\rm 1\mskip-4mu l} {\rm 1\mskip-4mu l}
{\rm 1\mskip-4.5mu l} {\rm 1\mskip-5mu l}}}
\usepackage[psamsfonts]{amssymb}
\begin{document}

\title{%
Doubly Special Relativity theories as different bases of
$\kappa$--Poincar\'e algebra}
\author{ J.\ Kowalski--Glikman\thanks{e-mail
address jurekk@ift.uni.wroc.pl;\ \ Research  partially supported
by the    KBN grant 5PO3B05620.}~~and S.\ Nowak\thanks{e-mail
address pantera@ift.uni.wroc.pl}\\ Institute for Theoretical
Physics\\ University of Wroc\l{}aw\\ Pl.\ Maxa Borna 9\\
Pl--50-204 Wroc\l{}aw, Poland} \maketitle

\begin{abstract}
Doubly Special Relativity (DSR) theory is a theory with two
observer-independent scales, of velocity and mass  (or length).
Such a theory has been proposed by Amelino--Camelia as a kinematic
structure which may underline quantum theory of relativity.
Recently another theory of this kind has been proposed by Magueijo
and Smolin. In this paper we show that both these theories can be
understood as particular bases of the $\kappa$--Poincar\'e theory
based on quantum (Hopf) algebra. This observation makes it
possible to construct the space-time sector of Magueijo and Smolin
DSR. We  also show how this construction can be extended to the
whole class of DSRs. It turns out that for all such theories the
structure of space-time commutators is the same. This results lead
us to the claim that physical predictions of properly defined DSR
theory should be independent of the choice of basis.
\newline

\end{abstract}

\clearpage

\section{Introduction}

About a year ago, in two seminal papers \cite{gac1}, \cite{gac2}
(see also \cite{gacnew})  G.~Amelino--Camelia proposed a theory
with two observer-independent kinematical scales: of velocity $c$
and of mass $\kappa$\footnote{In his papers Amelino--Camelia  uses
the scale of length $\lambda$ instead of the scale of mass
$\kappa$, however since the construction presented there describes
the energy--momentum sector of the theory, it seems more natural
to use the scale of mass. It should be also noted that it was the
scale of mass (more precisely of momentum, $\kappa c$), and not of
length, that has been shown explicitly to be
observer-independent.}. There are two major motivations for such
extension of special relativity. The first stems from the quest
for quantum gravity in which the Planck length is supposed to play
a fundamental role. For example, in loop quantum gravity the area
and volume operators have discrete spectra, with minimal value
proportional to the square and cube of Planck length, respectively
\cite{RovSmo}. The basic observation is that if one regards the
Planck length as a fundamental, intrinsic characteristic of the
space--time structure, this length should be the same for all
observers and thus one immediately finds himself in conflict with
FitzGerald--Lorentz contraction. The only solution of this problem
is to modify the principles of special relativity so as to
incorporate the existence of this second observer-independent
scale. It was indicated \cite{gac1}, \cite{gac2} that a possible
candidate for such a theory might be a theory based on deformed
Poincar\'e symmetry, for example the $\kappa$--Poincar\'e theory
\cite{lunoruto, maru, luruto, luruza, luno}. This claim was then
proved to be correct: it has been explicitly shown in the papers
\cite{jkgminl}, \cite{rbgacjkg} that the $\kappa$--Poincar\'e
theory in the so-called bicrossproduct basis indeed predicts the
existence of an observer-independent maximal mass. The second
motivation comes from some puzzling observations of ultra-high
energy cosmic rays, whose existence seems to contradict the
standard understanding of astrophysical process like $e^+$-- $e^-$
production in $\gamma\gamma$ collisions and the photopion
production in the scattering of high energy protons with soft
photons. It turns out that both these effects can be explained if
one assumes that the threshold condition becomes deformed,
possibly as a result of existence of a new fundamental length (or
mass) scale (see \cite{gacpir} and references therein). It is
exciting to note that it is likely that this purely quantum
gravitational effect will be a subject of numerous experimental
tests in the near future.

Most of the works on the relativity theory with two observed
independent kinematical scales,   dubbed ``Doubly Special
Relativity'' (DSR), has been done  in the framework of (or
strongly motivated by) algebraic construction based on the quantum
(Hopf) $\kappa$--Poincar\'e algebra, being a deformation of the
standard Poincar\'e algebra of special relativity. Recently
however Magueijo and Smolin \cite{JoaoLee} have proposed a
seemingly completely different DSR. The relation between these two
theories was thoroughly analyzed in \cite{gacandstud}. The
existence of two DSR's raises an obvious question  how many
theories of this kind may exist. In this paper we show that from
the quantum algebraic point of view both above mentioned DSR
theories are in fact completely equivalent, and might be
considered as  representations of the same $\kappa$-Poincar\'e
algebra in different bases. Moreover, one can easily construct
different yet representations of this algebra, each of whose would
correspond to a different DSR theory. Therefore, in what follows
we would use the notion of different basis instead of different
DSR.

The plan of this paper is the following. In the next section we
present three bases of $\kappa$-Poincar\'e algebra:  the
bicrossproduct one, lying behind the first DSR construction, the
Magueijo--Smolin one, and the classical one, whose algebraic
sectors is identical with the standard Poincar\'e algebra, and we
derive the transformations relating them. In section 3 we make use
of the co-algebraic sector of these bases to derive the space-time
non-commutative structure and to extend it to the whole of the
phase space. Section 4 is devoted to physical interpretation of a
picture resulting from mathematical constructions presented in
sections 2 and 3.

\section{Lorentz algebra and energy--momentum sector}

One of the main assumption in construction of DSR is that Lorentz
subalgebra of the $\kappa$-Poincar\'e algebra is  not to be
deformed. This assumption is motivated by the fact that one wants
to work with Lorentz structure that integrates to a group, and not
to a quasigroup. It restricts the possible choices of bases
severely (note that this postulate is not satisfied by the
so-called standard basis of the $\kappa$-Poincar\'e algebra
\cite{lunoruto}.) We therefore assume that the three rotation
generators $M_i$ and three boost generators $N_i$ satisfy
\begin{equation}\label{1}
 [M_i, M_j] = i\, \epsilon_{ijk} M_k, \quad [M_i, N_j] = i\, \epsilon_{ijk} N_k,
 \quad [N_i, N_j] = i\, \epsilon_{ijk} M_k.
\end{equation}
One  also assumes that the action of rotations is not deformed and
that generators of momenta commute. Taking these  postulates as a
starting point, we can define the (deformed) action of the Lorentz
algebra on energy--momentum sector. Our starting point would be
the bicrossproduct basis in which the resulting algebra is a
quantum algebra, i.e., in addition to the algebra of commutators
(which is usually non-linear)  it possess additional structures:
co-product $\Delta$ and antipode $S$. However one should remember
that quantum algebra structure is built on an enveloping algebra,
which means that one is entitled to make any transformations among
generators (and not only the linear one as in the case of Lie
algebras.) This leaves, of course, a lot of freedom in the choice
of energy and momentum generators.

\subsection{The bicrossproduct basis}

Since, as said above the action of rotations is standard it is
sufficient to write down only the commutators  of deformed boost
generators with momenta. One gets \cite{maru}
\begin{equation}\label{2}
   \left[N_{i}, p_{j}\right] = i\,  \delta_{ij}
 \left( {\kappa\over 2} \left(
 1 -e^{-2{p_{0}/ \kappa}}
\right) + {1\over 2\kappa} \vec{p}\,{}^{ 2}\, \right) - i\,
{1\over \kappa} p_{i}p_{j} ,
\end{equation}
and
\begin{equation}\label{3}
  \left[N_{i},p_{0}\right] = i\, p_{i}.
\end{equation}
One can easily check that the first Casimir operator of the
algebra (\ref{1}) -- (\ref{3})  reads\footnote{It turns out that
this form of the Casimir, which was used in all the papers devoted
to $\kappa$-Poincar\'e algebra and its application is not
physical, see section 4 for detailed discussion.}
\begin{equation}\label{6}
 m^2 = \left(2\kappa \sinh \left(\frac{p_0}{2\kappa}\right)\right)^2 - \vec{p}\,{}^2\, e^{p_0/\kappa}.
\end{equation}
It follows that for positive $\kappa$ the three-momentum is
bounded from  above $\vec{p}\,{}^2 \leq \kappa^2$ and the maximal
value of momentum corresponds to infinite energy \cite{jkgminl},
\cite{rbgacjkg}.
\newline

The quantum algebra (Hopf) structure in this basis is provided by the following co-products

\begin{eqnarray}
\displaystyle
 && \Delta(M_{i}) = M_{i}\otimes \bbbone + \bbbone
\otimes M_{i}\, ,
\cr\cr
\displaystyle
&& \Delta(N_{i}) = N_{i}\otimes \bbbone  +
e^{-{p_{0}/ \kappa}}\otimes N_{i} + {1\over \kappa}
\epsilon_{ijk}p_{j}\otimes M_{k}\, ,
\cr\cr
\displaystyle
&& \Delta(p_{i}) = p_{i}\otimes \bbbone +
e^{-{p_{0}/ \kappa}} \otimes p_{i}\, ,
\cr\cr
\displaystyle
&& \Delta(p_{0}) = p_{0}\otimes \bbbone +  \bbbone \otimes p_{0}\, ,
\label{4}
\end{eqnarray}
and the antipodes
\begin{eqnarray} 
\displaystyle
 && S(M_{i}) = - M_{i}\, ,
\cr\cr \displaystyle && S(N_{i}) = -e^{{p_{0}/ \kappa}}\,
N_{i}+\frac1\kappa\, \epsilon_{ijk}e^{{p_{0}/ \kappa}}\,p_j M_k \,
, \cr\cr \displaystyle && S(p_{0}) = -P_{0}\, , \cr\cr
\displaystyle && S(P_{i}) = -e^{{p_{0}/ \kappa}}\, P_{i}.
\label{4a}
\end{eqnarray}

Taking these formulas as a starting point, we can now turn to analysis of another bases.

\subsection{Magueijo--Smolin basis}

In the recent paper \cite{JoaoLee} Magueijo and Smolin proposed
another DSR theory, whose boost  generators were constructed as a
linear combination of the standard Lorentz generators and the
generator of dilatation (but in such a way that the algebra
(\ref{2}) holds.) In this basis the commutators of four-momenta
$P_\mu$ and boosts have the following form
\begin{equation}\label{7}
   \left[N_{i}, P_{j}\right] =  i\left( \delta_{ij}P_0 -
  {1\over \kappa} P_{i}P_{j} \right),
\end{equation}
and
\begin{equation}\label{8}
  \left[N_{i},P_{0}\right] = i\, \left( 1 - {P_0\over \kappa}\right)\,P_{i}.
\end{equation}
It is easy to check that the Casimir for this algebra has the form
\begin{equation}\label{9}
 M^2 = \frac{P_{0}^2 - \vec{P}{}^2}{\left(1- \frac{P_0}\kappa\right)^2}.
\end{equation}
The question arises as to if this basis is equivalent to the
bicrossproduct basis above. The answer is  affirmative, indeed one
easily checks that the relation between variables $P_\mu$ and
$p_\mu$ is given by
\begin{equation}\label{10}
p_{i} = P_{i}
\end{equation}
\begin{equation}\label{11}
p_0 = - \frac\kappa2\log\left(1 - \frac{2P_0}{\kappa} + \frac{\vec{P}{}^2}{\kappa^2}\right),
\quad  P_0 = \frac\kappa2\,\left(1 -  e^{-2p_0/\kappa} + \frac{\vec{p}\,{}^2}{\kappa^2}\right).
\end{equation}
Let us note by passing that  the  formula above shows that the
maximal momentum in the bicrossproduct basis  ($\vec{p}\,{}^2 =
\kappa^2$, $p_0=\infty$) corresponds to the maximal energy in the
Magueijo--Smolin basis, $P_0 =\kappa$.

 Using  formulas above one can without difficulty promote this algebra to the quantum algebra. This
 amounts only in using the homomorphisms (\ref{10}), (\ref{11}) to define the new co-products and antipodes. They read
\begin{equation}\label{6s}
  \triangle(P_{i})=P_{i} \otimes 1 + \left(1 - \frac{2P_{0}}{\kappa} +
   \frac{\vec{P}^{2}}{\kappa^{2}}\right)^{{1}/{2}} \otimes P_{i}
\end{equation}
$$
  \triangle(P_{0})=P_{0} \otimes 1 + 1 \otimes P_{0} -
  \frac{2}{\kappa}P_{0} \otimes P_{0} + \frac{1}{\kappa^{2}}
   \vec{P}^{2}\otimes P_{0} +
$$
\begin{equation}\label{7s}
   + \frac{1}{\kappa} \left(1 - \frac{2P_{0}}{\kappa} +
   \frac{\vec{P}^{2}}{\kappa^{2}}\right)^{{1}/{2}} \, \sum P_{i}\otimes P_{i}
\end{equation}
\begin{equation}\label{8s}
    S(P_{0})=\left(1 - \frac{2P_{0}}{\kappa} + \frac{\vec{P}^{2}}{\kappa^{2}}\right)^{-1}
  \left(\frac{\vec{P}^{2}}{2\kappa} - \frac{\kappa}{2}\right) + \frac{\kappa}{2}
\end{equation}
\begin{equation}\label{9s}
  S(P_{i})=-P_{i}\left(1 - \frac{2P_{0}}{\kappa} + \frac{\vec{P}^{2}}{\kappa^{2}}\right)^{-1/2}
\end{equation}

\subsection{The classical basis}

There is yet another basis which we will present here for
comparison (this basis was first described in \cite{maslanka},
\cite{kolumaso}, \cite{lukclas}, \cite{luruza}.) In this basis,
which we call the classical one, the boosts--momenta commutators
together with the Lorentz sector form the classical Poincar\'e
algebra, to wit
\begin{equation}\label{12}
   \left[N_{i}, {\cal P}_{j}\right] = i\, \delta_{ij}\, {\cal P}_0  ,\quad \left[N_{i}, {\cal P}_{0}\right] = i\,
   {\cal P}_i.
\end{equation}
The Casimir for this basis equals, of course the one of special relativity, to wit
\begin{equation}\label{12a}
 {\cal M}^2 = {\cal P}_0^2 - \vec{\cal P}\,{}^2
\end{equation}
The classical generators ${\cal P}_\mu$ are related to the bicrossproduct basis generators by the formulas
\begin{equation}\label{13}
 {\cal P}_{0} = \kappa \sinh\frac{p_0}\kappa + e^{p_0/\kappa}\, \frac{\vec{p}\, {}^2}{2\kappa},
\end{equation}
\begin{equation}\label{14}
 {\cal P}_{i}= e^{p_0/\kappa} \, p_i
\end{equation}
and one can easily compute the expression for co-product
$$
  \Delta({\cal P}_{0}) = \frac\kappa2\left(K \otimes K - K^{-1}\otimes K^{-1}\right)+$$
  \begin{equation}\label{15}
  + \frac1{2\kappa}\left(K^{-1} \vec{{\cal P}}{}^2 \otimes K +
  2K^{-1} {\cal P}_i \otimes {\cal P}_i + K^{-1}\otimes K^{-1}\vec{{\cal P}}{}^2\right),
\end{equation}
\begin{equation}\label{15a}
 \Delta({\cal P}_{i}) ={\cal P}_{i}\otimes K + \bbbone\otimes {\cal P}_{i}
\end{equation}
where
$$
K = e^{p_0/\kappa} = \frac1\kappa\, \left[ {\cal P}_{0} + \left(
{\cal P}_{0}^2 - \vec{{\cal P}}{}^2  +
\kappa^2\right)^{1/2}\right],
$$
and the antipode
\begin{equation}\label{15s1}
    S({\cal P}_{0})=-{\cal P}_{0} + \frac1\kappa\vec{{\cal
    P}}{}^2\, K^{-1}
\end{equation}
\begin{equation}\label{15s2}
  S({\cal P}_{i})=-{\cal P}_{i}\, K^{-1}
\end{equation}

\section{The space-time non-commutativity}

In the preceding section we investigated the energy--momentum
algebras. Now it is time to explain the  relevance of the
co-product (quantum) structure of these algebras. Briefly, this
structure makes it possible to extend an energy-momentum algebra
to the whole of a phase space, i.e., the space describing both
energy-momentum and space-time sectors in self consistent way. One
should stress that this is the only way to interconnect the
space-time and energy-momentum sectors in a systematic way. It
turns out that the co-algebra  of the energy-momentum sector is in
one to one correspondence with algebraic sector of space-time
algebra and vice versa. Observe that in order to construct such a
correspondence we need one more dimensionful parameter, which we
identify with the Planck constant (in the following we will use
the convention in which $\hbar=1$.) We will comment on this point
in the next section.

There is a  general procedure how to construct the space-time
commutator algebra from energy-momentum  co-algebra, which
consists of the following steps \cite{maru}, \cite{crossalg}:

\begin{enumerate}
\item One defines the bracket $<\star, \star>$ between momentum variables $p,q$ and position variables $x,y$ in
a natural way as follows
\begin{equation}\label{16a}
 <p_\mu, x_\nu> =  -i \eta_{\mu\nu}, \quad \eta_{\mu\nu} = \mbox{diag}(-1,1,1,1).
\end{equation}
\item This bracket is to be consistent with the co-product structure in the following sense
\begin{equation}\label{16b}
 <p, xy> = <p_{(1)}, x><p_{(2)}, y>, \quad <pq,x> =<p, x_{(1)}><q_{(2)}, x_{(2)}>,
\end{equation}
where we use the natural notation for co-product $$\Delta t = \sum
t_{(1)} \otimes t_{(2)}.$$ It should be also  noted that by
definition $$<\bbbone, \bbbone> =1.$$ One sees immediately that
the fact that momenta commute translates to the fact that
positions co-commute
\begin{equation}\label{16c}
  \Delta x_\mu = \bbbone \otimes x_\mu + x_\mu \otimes \bbbone.
\end{equation}
Then the first equation in (\ref{16b}) along with (\ref{16a}) can be used to deduce the form of the space-time commutators.
\item  It remains only to derive the cross relations between momenta and positions. These can be found
from the definition of the so-called Heisenberg double (see \cite{crossalg}) and read
\begin{equation}\label{16d}
 [p,x] =  x_{(1)}<p_{(1)}, x_{(2)}>p_{(2)}-xp
\end{equation}
\end{enumerate}

As an example let us perform these steps in the bicrossproduct basis \cite{maru}, \cite{crossalg}.
It follows from (\ref{16b}) that
$$
<p_i, x_0 x_j> = -\frac1\kappa\, \delta_{ij}, \quad <p_i,  x_jx_0>=0,
$$
from which one gets
\begin{equation}\label{17a}
[x_0, x_i] = -\frac{i}\kappa\, x_i.
\end{equation}
Let us now make use of (\ref{16d}) to get the standard relations
\begin{equation}\label{17b}
[p_0, x_0] = i, \quad [p_i, x_j] = -i \, \delta_{ij}.
\end{equation}
However it turns out that this algebra contains one more
non-vanishing commutator, namely
\begin{equation}\label{17c}
 [p_i, x_0] = -\frac{i}\kappa\, p_i.
\end{equation}
Of course, the algebra (\ref{17a}--\ref{17c}) satisfies the Jacobi identity.

\subsection{Magueijo--Smolin basis}

To find the non-commutative structure of space time in Magueijo--Smolin basis we start again with eq.~(\ref{16a})
$$
 <P_\mu, X_\nu> =  -i \eta_{\mu\nu}.
$$
Let us now turn to the next step, eq.~(\ref{16b}). It is easy to
see that the only terms in (\ref{6s}), \ref{7s}),  which are
relevant for our computations are the bilinear ones, so we can
write
$$
  \triangle(P_{i})=\bbbone \otimes P_{i} + P_{i} \otimes \bbbone {-\frac{1}{\kappa}}\, P_{0} \otimes P_{i} + \ldots
$$
$$
  \triangle(P_{0})= \bbbone \otimes P_{0} + P_{0} \otimes \bbbone -
  \frac{2}{\kappa}P_{0} \otimes P_{0} +  \frac{1}{\kappa}\sum P_{i}\otimes P_{i}+ \ldots
$$
It follows immediately that the only non-vanishing commutators in the position sector are
\begin{equation}\label{18}
  [X_0, X_i] = -\frac{i}\kappa\, X_i.
\end{equation}
Now we can use eq.~(\ref{16d}) to derive the form of the remaining
commutators. Since this computation is a bit tricky, let us
present the necessary steps.
$$
[P_0, X_i] = \sum_j \left<\frac{1}{\kappa} \sqrt{1 -
\frac{2P_{0}}{\kappa} +
   \frac{\vec{P}^{2}}{\kappa^{2}}} \,\,  P_{j}, X_i\right>\, P_j   +
X_i <\bbbone, \bbbone>\, P_0 - X_i P_0 =
$$
$$
=\frac{1}{\kappa}\,\sum_j <P_{j}, X_i>\, P_j
$$
(we made use of the fact that the only terms linear in momenta
have non-vanishing bracket with positions) from which it follows
immediately that
\begin{equation}\label{19b}
 [P_0, X_i] =  -\frac{i}\kappa P_i
\end{equation}
and by employing the same procedure we obtain the remaining
commutators
\begin{equation}\label{19a}
 [P_0, X_0] = i\left(1 - \frac{2P_0}\kappa \right)
\end{equation}
\begin{equation}\label{19c}
  [P_i, X_j] = -i \, \delta_{ij}
\end{equation}
\begin{equation}\label{19d}
 [P_i, X_0] = -\frac{i}\kappa\, P_i.
\end{equation}
Of course, as it is easy to check, the algebra above satisfies the
Jacobi identity.

\subsection{The classical basis}

We again start with the duality relation
$$
 <{\cal P}_\mu, {\cal X}_\nu> =  -i \eta_{\mu\nu},
$$
and to get the commutators in the position sector as above we take
 the part of the co-product up to the bilinear terms
$$
  \triangle({\cal P}_{i})=\bbbone \otimes {\cal P}_{i} + {\cal P}_{i} \otimes \bbbone +
  {\frac{2}{\kappa}}\, {\cal P}_{0} \otimes {\cal P}_{i} + {\frac{1}{\kappa}}\, {\cal P}_{i} \otimes {\cal P}_{0} + \ldots
$$
$$
  \triangle({\cal P}_{0})= \bbbone \otimes {\cal P}_{0} + {\cal P}_{0} \otimes \bbbone +
  \frac{1}{\kappa}\sum {\cal P}_{i}\otimes {\cal P}_{i}+ \ldots
$$
which leads again to
\begin{equation}\label{20}
  [{\cal X}_0, {\cal X}_i] = -\frac{i}\kappa\, {\cal X}_i.
\end{equation}
Then by employing the same method as in the preceding subsection
we find the space-time commutators
\begin{equation}\label{21a}
 [{\cal P}_0, {\cal X}_0] = \frac{i}2\left(K + K^{-1}- \frac1{\kappa^2}\vec{{\cal
    P}}{}^2\, K^{-1}\right)
\end{equation}
\begin{equation}\label{21b}
 [{\cal P}_0, {\cal X}_i] =  -\frac{i}\kappa {\cal P}_i
\end{equation}
\begin{equation}\label{21c}
  [{\cal P}_i, {\cal X}_j] = - i \, K\delta_{ij}
\end{equation}
\begin{equation}\label{21d}
 [{\cal P}_i, {\cal X}_0] = 0.
\end{equation}
\vspace{12pt}

Let us summarize the results of obtained so far. We found the
transformations in the energy--momentum sector relating the
bicrossproduct, Magueijo--Smolin, and classical bases of
$\kappa$-Poincar\,e algebra. Next we extended this construction to
the space-time sector. This examples strongly suggest that the
following two claims hold under mild and physically acceptable
assumption concerning the change of bases (one must assume that in
any basis  an algebra becomes the standard Poincar\,e algebra in
the leading order in $\kappa$ expansion)

(a) Any algebra
consisting of the undeformed Lorentz sector  with the standard
action of rotations and generic action of boosts on commuting
four-momenta can be equipped with the $\kappa$-Poincar\,e quantum
algebraic structure and

(b) This structure can be extended to the whole of the phase
space, and the non-commutative structure of the space-time sector
is always (i.e., in any basis) given by
\begin{equation}\label{22}
  [x_0, x_i] =-\frac{i}\kappa\, x_i.
\end{equation}
The precise formulation of these statements and their proof will
be presented in forthcoming paper.

\section{Physical interpretation}

Till now we have been dealing with mathematics, it is time
therefore to turn to physics. In view of remarks at the end of
preceding section the situation is as follows. Instead of having a
number of distinct DSR theories we have to do with a whole class
of such theories, which are related to each other by redefinition
of momenta
\begin{equation}\label{23}
  p_0 \rightarrow f(p_0, \vec{p}\, {}^2), \quad p_i \rightarrow g(p_0, \vec{p}\,
  {}^2)\, p_i
\end{equation}
 restricted only by the requirement that the action of
rotations is preserved, each of whose can be interpreted as being
a particular basis of $\kappa$-Poincar\,e quantum algebra. If we
are to base a physical theory on such a mathematical structure, we
have clearly two choices either to find some physical and/or
mathematical motivation to single out one particular basis, or to
define the physical theory in such a way that its physical
predictions are independent of the choice of a basis. This last
possibility reminds very much the postulate of general coordinate
invariance of general relativity and seems worth pursuing. From
this perspective the fact that one of the most fundamental
prediction of the theory, namely the non-commutativity in the
space-time sector (\ref{22}) is invariant under the change of
basis is very encouraging. However one  should note that in such a
case, the ``momentum'' variables do not have a direct physical
meaning. This does not seem very surprising after all. Indeed, in
special relativity as well as in non-relativistic mechanics anergy
and momentum are defined so as to be conserved in physical process
and this conservation is regarded as one of the most important
physical properties of nature. On the other hand energy/momentum
conservation is directly related to the homogeneity of space-time.
Here we have to do with non-commutative space-time and the
presence of length scale strongly suggest that space-time is
homogeneous only at scales much larger than this scale. It is not
clear how to define the energy/momentum conservation (and thus to
answer the question what is the {\em physical} energy and
momentum) for a non-commutative space-time, but it is certain that
any such definition must include information concerning the
space-time non-commutativity.

Let us turn to another point. It is clear that one of the most
important physical characteristic of a  particle is its rest mass.
This mass should be, as in the special relativity and
non-relativistic mechanics equal to the Casimir operator of the
theory at hands. It is reasonable therefore to adopt the following
definition of physical rest mass
\begin{equation}\label{24}
  \frac{1}{m_{0}}=\lim_{p\rightarrow0}\frac{1}{p}\frac{dE}{dp}, \quad E \equiv p_0
\end{equation}
 It is easy to check that this definition
works in special relativity, where $m^2 = E^2 - p^2$ and in
non-relativistic mechanics, where $m = p^2/2E$. Let us now check
whether the Casimirs in bases discussed above are  physical
masses. This is clearly the case in the classical basis. In the
Magueijo--Smolin basis one gets
$$
{M_0^2} = \frac{E^2}{\left(1-\frac E\kappa\right)^2} = M^2
$$
and thus the Magueijo--Smolin Casimir (\ref{9}) indeed equals the physical rest mass.

The situation in the bicrossproduct basis is different, however. One easily finds that in this basis
\begin{equation}\label{25}
{m_{0}}^2=\frac{\kappa^2}{4}\, \left(1-e^{-2p_0/\kappa}\right)^2
=\frac{\kappa^2}{4}\, \left(1-\left(-\frac{m}{2\kappa} +
\sqrt{\frac{m^2}{4\kappa^2} +1}\right)^4\right)^2,
\end{equation}
where $m^2$ is given be eq.~(\ref{6}). This result is of
importance because it suggests that the form of dispersion
relation used in the literature so far might be incorrect. This
suggests in particular that one should rethink the status of
$\kappa$-Poincar\,e as a possible explanation of the threshold
anomalies in cosmic rays astrophysics (cf.~\cite{gacpir}.) We will
return to this question in separate paper.

It is also interesting to note the following relation between
physical masses in bicrossproduct ($m_0$), Magueijo--Smolin
($M_0$), and classical (${\cal M}_0$)bases
\begin{equation}\label{26}
  \frac1{M_0} = \frac1{m_0} - \frac1\kappa, \quad \frac1{{\cal M}^2_0} = \frac1{M^2_0} -
  \frac1{\kappa^2}.
\end{equation}
This suggests that an appropriately defined physical mass might be
invariant under change of basis and therefore a candidate for the
second ,,observable'' of DSR if this theory could be consistently
constructed in the basis-independent way.

Another point to be stressed is the following. In the DSR theory
proposed in \cite{gac1}, \cite{gac2} one have to do with two
scales of velocity and of mass. It should be noted that here we
have three scales (in our notation two of them $c$ and $\hbar$
were put equal  $1$): of speed $c$, which relates time and space
components of physical quantities, of mass $\kappa$, which governs
the deformation of the energy--momentum sector, and, finally, the
Planck constant $\hbar$, which makes it possible to relate the
energy--momentum and space--time sectors. This means that in fact
we have to do with Triply not Doubly Special Relativity. Observe
that these three scales must be present in the theory to make the
construction presented above possible. Of course, the question
arises as to what is the physical status of these scale: are they
observer-independent quantities or just coupling constants. To
answer this question one must find out what is an operational
definition of each of them.

\section{Conclusions }

Let us summarize the basic results of this paper. We showed that
DSR theories analyzed so far can be understood as different bases
of the $\kappa$-Poincar\,e theory. This observation made it
possible to construct the space-time sector of these theories.
This construction, in turn, led us to the claim that the DSR
theory is perhaps based on principle similar to diffeomorphism
invariance of general relativity, namely that the physical
predictions of DSR should be independent of the choice of basis.

\section*{Acknowledgement}

We would like to thank J.~Lukierski for useful comments.

\end{document}